\documentclass[preprints,article,accept,moreauthors]{mdpi} 

\firstpage{1} 
\pubvolume{1}
\issuenum{1}
\articlenumber{0}
\pubyear{2023}
\copyrightyear{2023}
\datereceived{} 
\dateaccepted{} 
\datepublished{} 
\hreflink{https://doi.org/} 
\pdfoutput=1

\Title{Gas and stars in the Teacup quasar looking with the 6-m telescope}

\TitleCitation{Gas and stars in the Teacup}

\newcommand{\SII}{[S~{\sc ii}]}
\newcommand{\OIII}{[O~{\sc iii}]}
\newcommand{\NII}{[N~{\sc ii}]}
\newcommand{\HII}{H~{\sc ii}}

\newcommand{\Ha}{H$\alpha$}
\newcommand{\HaHb}{$I$(H$\alpha)/I$(H$\beta$)}
\newcommand{\Hb}{H$\beta$}
\newcommand{\kms}{\,\mbox{km}\,\mbox{s}^{-1}}
\newcommand{\ergs}{\,\mbox{erg}\,\mbox{s}^{-1}}

\newcommand{\NIIHa}{$I$([N\,II])/$I$(H$\alpha$)}
\newcommand{\OIIIHb}{$I$([O\,III]\,$\lambda$\,5007)/$I$(H$\beta$)}
\newcommand{\apj}{ApJ}
\newcommand{\apjl}{ApJL}
\newcommand{\aap}{A\&A}

\newcommand{\aj}{AJ}
\newcommand{\mnras}{MNRAS}

\newcommand{\pasp}{PASP}
\newcommand{\apjs}{ApJS}
\def\red{}

\Author{Alexei V. Moiseev $^{1,\dagger,\ddagger}$*\orcidA{} and Alina I.  Ikhsanova $^{2,\ddagger}$}

\AuthorNames{Alexei V. Moiseev  and Alina I. Ikhsanova}

\AuthorCitation{Alexei V. Moiseev  and Alina I. Ikhsanova}

\address{%
$^{1}$ \quad Special Astrophysical Observatory, Russian Academy of Sciences, Nizhny Arkhyz 369167, Russia; moisav@gmail.com\\
$^{2}$ \quad  Dipartimento di Fisica e Astronomia “G. Galilei”, Università di Padova, vicolo dell’Osservatorio 3, 35122 Padova, Italy}

\corres{Correspondence: moisav@gmail.com;  (A.V.)}

\abstract{New results on the radio-quiet type 2 quasar,  known as the Teacup galaxy (SDSSJ1430+1339), based on the long-slit and 3D spectroscopic data obtained at the Russian 6-m telescope, are presented. The ionized gas giant nebula, which  extends up to $r=56$ kpc in the \OIII{} emission line, was mapped with the scanning Fabry--Perot interferometer. The direct estimation of the emission line ratios confirmed that the giant nebula is ionized by the AGN.  Stars in the inner $r<5$ kpc are significantly younger than the outer host galaxy and have a solar metallicity. The central starburst age ($\sim$1 Gyr) agrees with possible ages for the galactic merger events and the previous episode of the quasar  outflow produced two symmetric arcs visible in the \OIII{} emission  at the distances  $r=$50--55 kpc.  The ionized gas velocity field can  be fitted by the model of a circular rotating disk significantly inclined or even polar to the stellar host galaxy.
}


\begin{document}

\section{Introduction} 
\label{sec:intro}

Extended Emission-Line Regions (EELRs) detected in outskirts of some active galaxies are considered as a result of Active Galactic Nuclei (AGN) feedback in the radiative  and kinetic (jet and wind outflows) forms. The Seyfert galaxy Mrk 6 gives us a nearby, recent and  a very typical example of EELR related with AGN ionization cone observed outside of stellar host up to a projected distance 40 kpc \citep{Smirnova2018MNRAS.481.4542S}. Study of such structures allows us to better  understand both the history of nuclear radiative output on the scales 0.01-0.1 Myr and the distribution of intergalactic medium \citep[][and references therein]{Morganti2017NatAs...1..596M,Knese2020MNRAS.496.1035K,Keel2022MNRAS.510.4608K}.

One of the  largest EELR among low-redshift radio-quiet AGN was recently found around  Teacup galaxy (SDSS J143029.88+133912.0). The galaxy was discovered by volunteers of the Galaxy Zoo project, its EELR was confirmed in follow-up spectroscopic observations  \citep{Gagne2011AAS...21714212G}. This type-2 quasar was nicknamed due to the morphology of ionized gas bubbles that extend 10 kpc of the  galactic centre in the shape of  `handle'.
\citet{Keel2017ApJ...835..256K} considered  Teacup as a fading AGN,  whereas X-ray data have indicated that it is possible that no fading is required \citep{Lansbury2018ApJ...856L...1L}.

The  Hubble Space Telescope (HST) imaging shows that the Teacup is a bulge-dominated galaxy, with a shell-like structure  and a tidal tail, which have been interpreted  as an indicative of  merger 1--2 Gyr ago with a cold disk system and 1/10 mass ratio \citep{Keel2015AJ....149..155K}. Another interesting feature of this system is the presence of a giant outflow generated by small-scale radio jets and/or  quasar winds (\citet{Harrison2015ApJ...800...45H},\citep{Keel2017ApJ...835..256K}). This outflow appears to be  responsible for the bubble-like morphology of the galaxy. The ionized gas kinematics of the outflow in the inner 15 kpc were studied several times using long-slit and 3D-spectroscopic technique \citep{Keel2012MNRAS.420..878K,Gagne2014ApJ...792...72G,Keel2017ApJ...835..256K, Ramos2017MNRAS.470..964R}. However, the stellar kinematics of the host galaxy is still unknown.

The long-slit Gran Telescopio Canarias (GTC) optical spectroscopy by  \citet{Villar2018MNRAS.474.2302V} reveals that the ionizied gas around Teacup  extending up to 50 kpc in the \Ha{} emission line. This  giant nebula has been considered by the authors as reservoir of the circumgalactic medium populated by tidal debris produced by galactic merger events. The external gas  is most likely  photoionized by the nuclear radiation is significantly dynamical cold comparing with the inner 10--20 kpc region. The recent GTC deep image in the \Ha{} emission line clearly demonstrates that the giant Teacup EELR ($155\times87$ kpc in a total size) elongates  in the same direction with the main axis of the inner bubble and radio jet  \citep{Villar2021}. Several arcs and emission knots are visible up to $\sim56$ kpc from the AGN. However, details of  gas rotation pattern were still unknown \citep[][proposed that the gas maybe settled in a giant rotating disk]{Villar2018MNRAS.474.2302V}.

The giant nebula is most likely  photoionized by the nuclear radiation, however to construct their diagnostic diagrams  \citet{Villar2018MNRAS.474.2302V} accepted the  \OIIIHb{}  flux ratio similar with the value obtained early for the internal regions, due to  the lack of a green lines in the GTC spectrum.

In this work we present results of new long--slit and 3D spectroscopic observations   at the 6-m telescope of the Special Astrophysical Observatory of the Russian Academy of Sciences (SAO RAS) performed to solve the  above-mentioned puzzles of the Teacup galaxy:   properties of the host stellar population, the   structure and the velocity field  of the extended nebula, including direct estimation of its ionization state. Following \citep{Villar2018MNRAS.474.2302V}, we accepted  Teacup redshift $z = 0.085$ that corresponds to the  distance    $360$ Mpc ($H_0$ = 71 kms$^{-1}$ Mpc$^{-1}$) and a spatial scale    1.58 kpc arcsec$^{-1}$.

\begin{figure*}
  \centering
\makebox[\textwidth]{
 \includegraphics[width=0.53\textwidth]{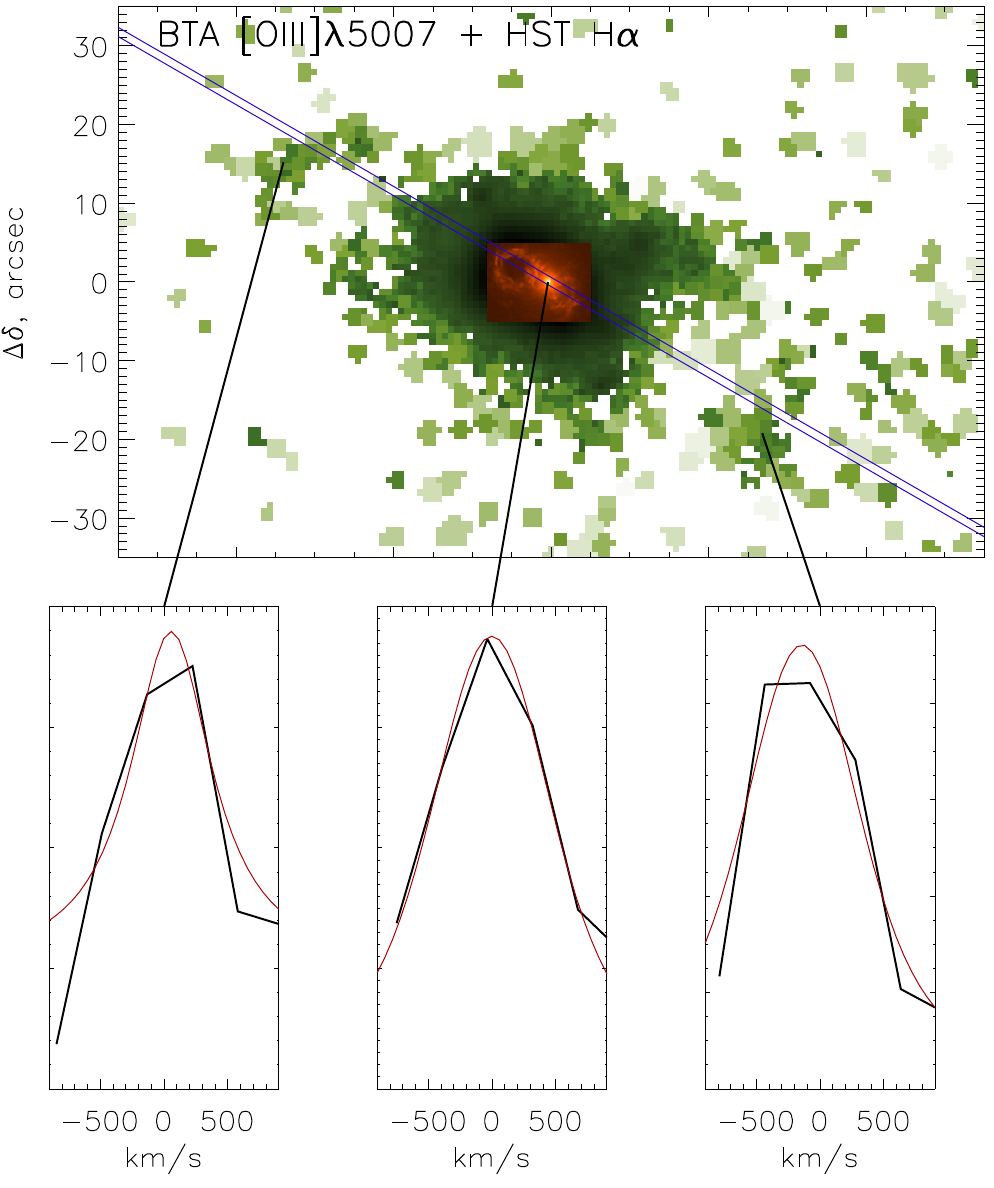}
\parbox[b]{0.47\textwidth}{
\includegraphics[width=0.47\textwidth]{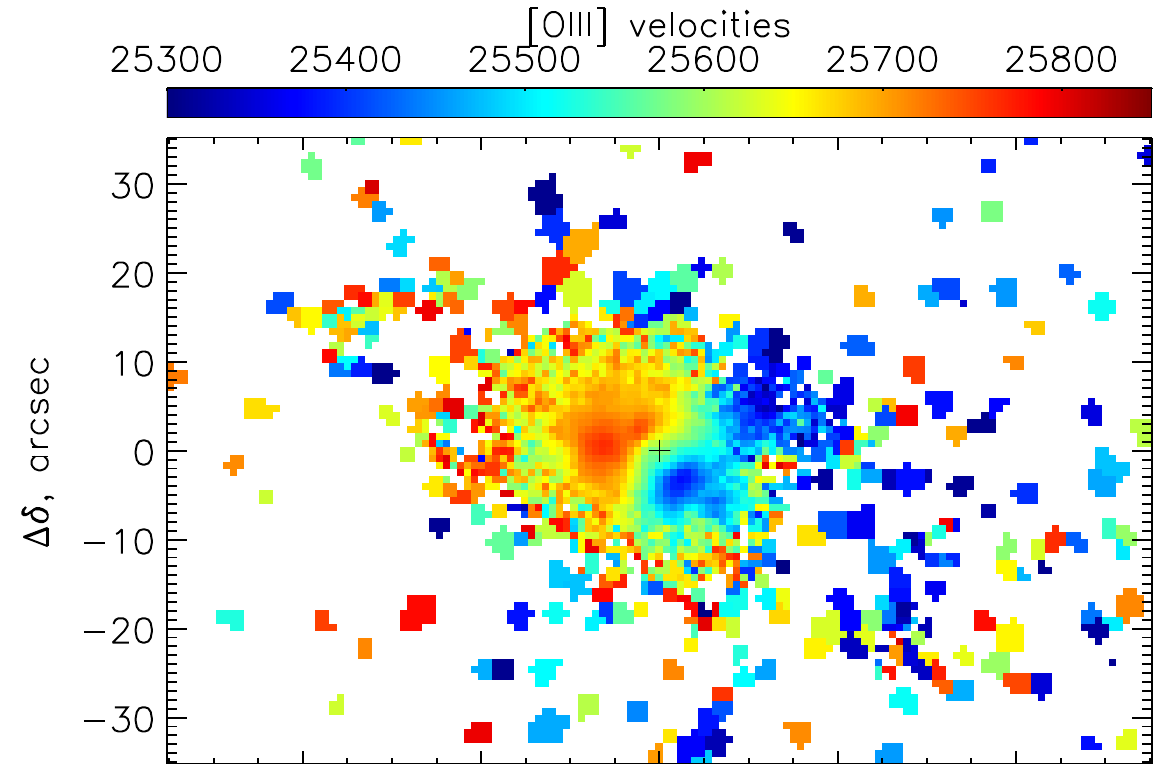}
\includegraphics[width=0.47\textwidth]{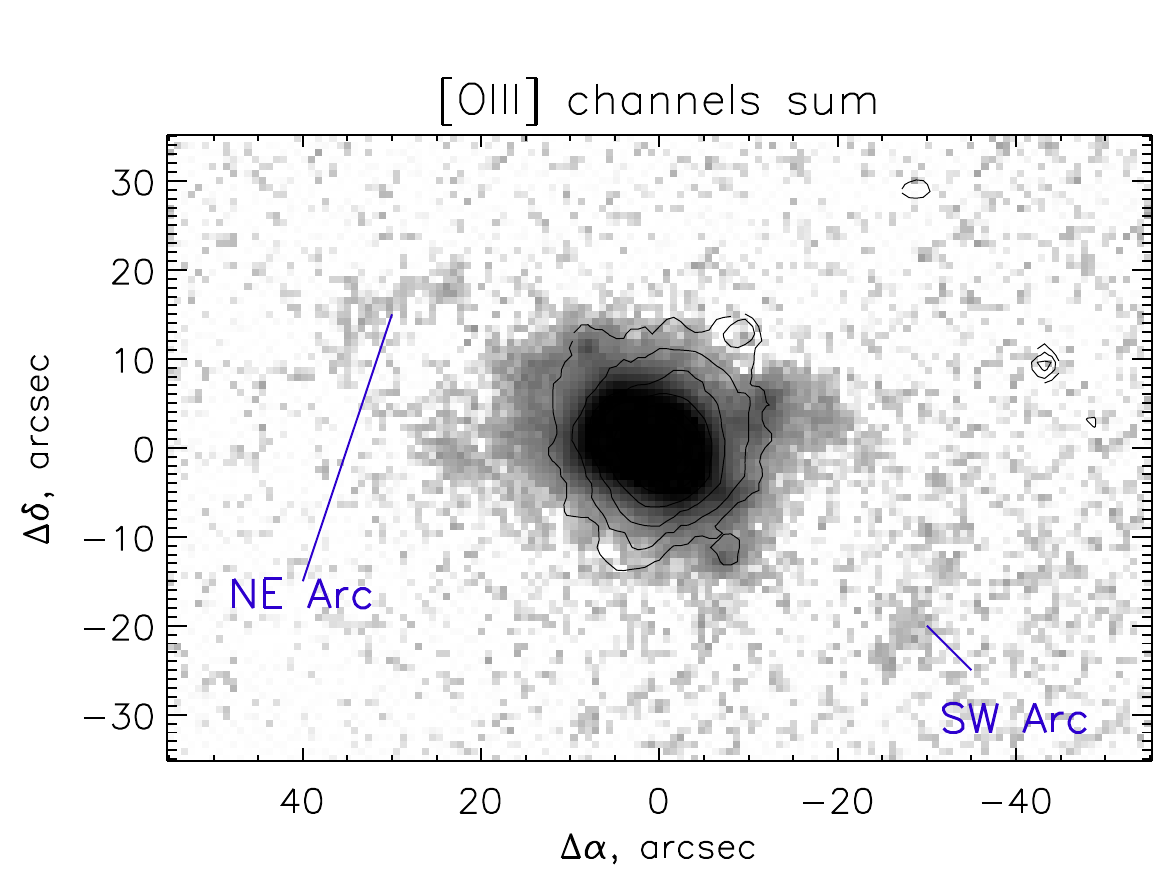}
 }
 }
\caption{The left panel: the \OIII{} emission  map taken with FPI after Voronoi tessellation binning (the green scale) combined with ACS HST \Ha{} image~\cite{Keel2015AJ....149..155K} of the central bubble region (in the red scale). The purple lines mark the position of the spectrograph slit ($1''$ in the width). The bottom plots show the emission-line profiles in the selected regions, the   Voigt fitting is marked by the red. The right    panels display the ionized gas velocity field (right top)  and the sum of   three channels centered at the  emission line in  the original $0.78''$ pixel scale with the   isophotes of the image in the channel correspond to the stellar continuum. The most distance emission filaments are marked: the NE Arc (firstly mapped in~Ref. \cite{Villar2021}) and SW Arc. 
}\label{fig1}
   \end{figure*}

\section{Observations and data analysis}
  
The observations were carried out at the prime focus of the SAO RAS  6m telescope with the SCORPIO-2   multi-mode focal reducer\citep{AfanasievMoiseev2011}. The detector was CCD $2K\times4.5K$ E2V 42-90 . The observations in the  long-slit mode  performed with the  $6'\times1''$ slit and the scale $0.36''/$px, other parameters are given in the Tab.~\ref{tab:spec_data} ($T_{exp}$ -- total exposure, $\theta$ -- mean seeing value, $\Delta\lambda$ and $\delta\lambda$ are  the spectral range and resolution). The position angle $PA=60^\circ$ corresponds to the major axis of the nebula (Fig.~\ref{fig1}). The initial data reduction were performed in a standard way as it was described in our previous papers \citep[e.g.][]{Egorov2018MNRAS.478.3386E}.  

The 3D-spectroscopy mapping in the \OIII$\lambda5007$ emission line was carried out in the scanning Fabry-Perot interferometer (FPI)  mode of SCORPIO-2 with the same low-resolution FPI  that is usually used  for the tunable filter imaging in the  MaNGaL device \citep{Mangal}. During the observations we subsequently obtained narrow band (the bandwidth $~\sim13$\AA) images   with different central wavelength: 5 frames spanned the spectral range around the redshifted \OIII$\lambda5007$ line with the step $5.9$\AA{} and one frame in continuum at the central wavelength shifted on 24\AA{}  from this emission line.  The field of view at the new detector CCD $2K\times4K$ E2V 261-80 was $6.8'$ sampled with a pixel scale  $0.78''$ and $0.39''$ at the nights 2020 Apr 19 and Apr 24 respectively, other parameters are given in Tab.~\ref{tab:spec_data}.  

\begin{table}[H]
 \caption{Log of SCORPIO-2/6m telescope observations
\label{tab:spec_data}}
		\newcolumntype{C}{>{\centering\arraybackslash}X}
		\begin{tabularx}{\textwidth}{llllll}
		\toprule
 	Mode     & Date of obs.    & $\mathrm{T_{exp}}$, s  & $\theta$, $''$   & $\Delta\lambda$, \AA & $\delta\lambda$, \AA  \\ 
		\midrule
	Long-slit & 2018 Feb 11 & $4\times1200$ & 2.2 & {3500--7220} & 5 \\
	FPI         & 2020 Apr 19       & $41\times90$  & 3.2 & 5410--5470  & 13 \\
	FPI         & 2020 Apr 24       & $50\times90$  & 1.6 & 5410--5470   & 13 \\

			\bottomrule
		\end{tabularx}
\end{table}

The preliminary data reduction (bias, flat-field correction, combining   individual 90 s exposures  with cosmic-ray hits cleaning) was performed with IDL-based software as described in \citet{Mangal} for the MaNGaL  scanning mode. The air-glow lines subtraction, photometric correction  and phase-shift wavelength calibration were done with our software for SCORPIO-2/FPI data reduction \citep[see description and references in ][]{Moiseev2021AstBu..76..316M}. The data obtained during two nights were merged  in the single data cube containing low-resolution  \OIII{} spectra in each pixel with the size $0.78''$. The total exposure  was about 2.4 h. The surface  brightness of the most faint detected emission filaments is about  $2\times10^{-18}\ergs\,\mbox{cm}^{-2}\,\mbox{arcsec}^{-2}$.

In order to improve signal-to-noise (S/N) ratio  in the  spectra of the faint outer region the Voronoi tessellation was used \citep{Cappellari2003MNRAS.342..345C}. The emission line in the binned data cube was fitted with  Voigt function that provides a good approximation of the observed FPI spectra \citep{Moiseev2008AstBu..63..181M}. The example of emission line profiles together   with the \OIII{} flux maps (in the binned and original resolution) and line-of-sight velocity field is shown in Fig.~\ref{fig1}.

\section{The stellar population properties and kinematics}
\label{sec:stars}

\begin{figure}[H]
\includegraphics[width=1\textwidth]{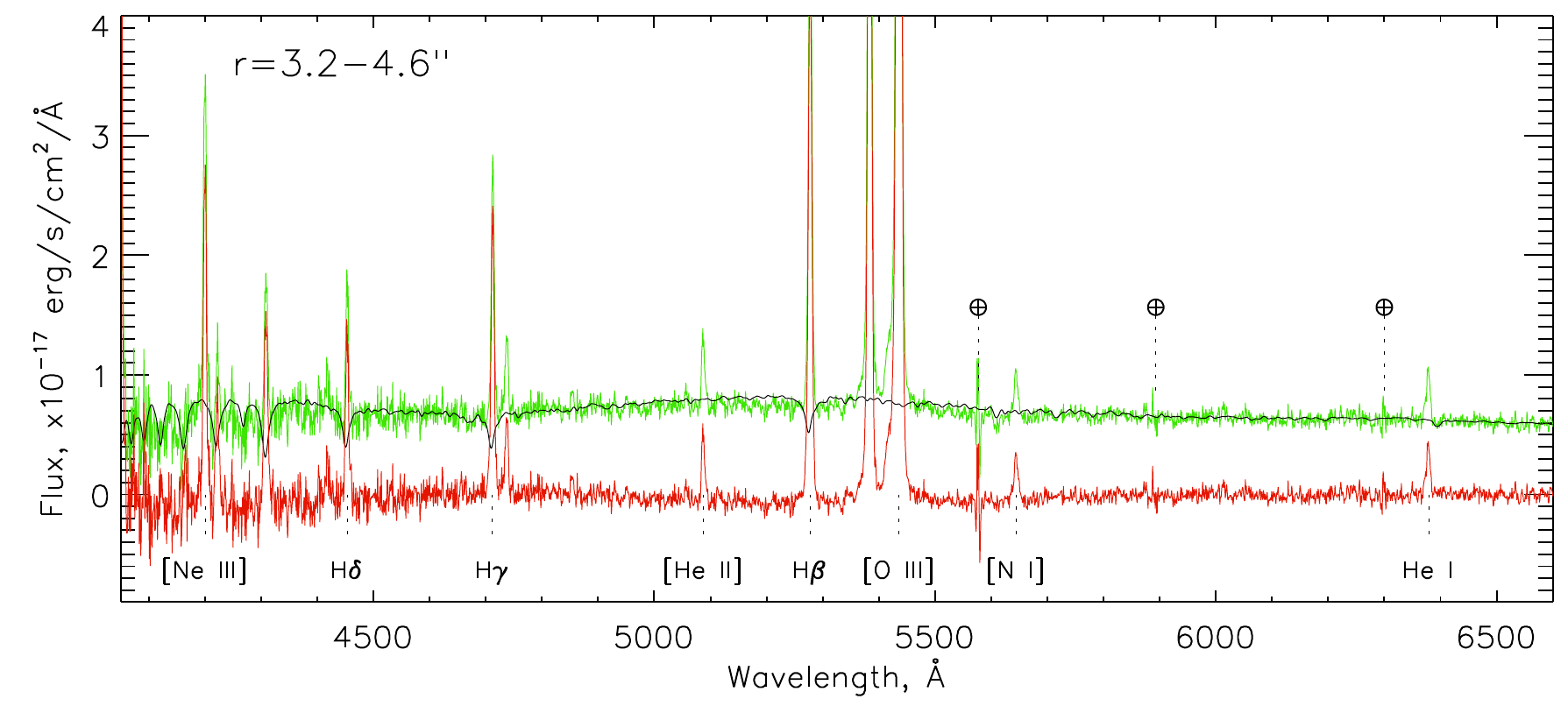}
\caption{\red{pPXF analisys of the observed spectrum (the green line) in the radial bin corresponded to distance from the Teacup nucleus  r=$3.2-4.6''$. The stellar population model is shown in  black, the red spectrum is a residual after model subtraction. The main ionized gas emission lines and position of the airglow lines are labeled. 
}}
\label{fig:pPXF}
  \end{figure}

The observed long-slit spectra contain the combination of the ionized gas emission and stellar  absorption lines. We used penalized pixel-fitting (pPXF) method to fit a stellar population  spectrum \citep{Cappellari2017MNRAS.466..798C}, using MILES stellar spectral library by \citet{Vazdekis2010MNRAS.404.1639V}   covering the range 3525--7500\AA{} with a twice higher spectral resolution than in the  SCORPIO-2 data. In order to improve the signal-to-noise ratio, the spectra used
for the pPXF analysis were binned along the slit using the step exponentially increasing with   radius: from 2 px ($0.7''$) bins in the nucleus up to  10 px ($3.6''$)  at the distances $r\sim10''$ from the center. \red{The example of the observed Teacup spectrum fitted by pPXF model is shown in Fig.~\ref{fig:pPXF}}.

\begin{figure}[H]
\includegraphics[width=1\textwidth]{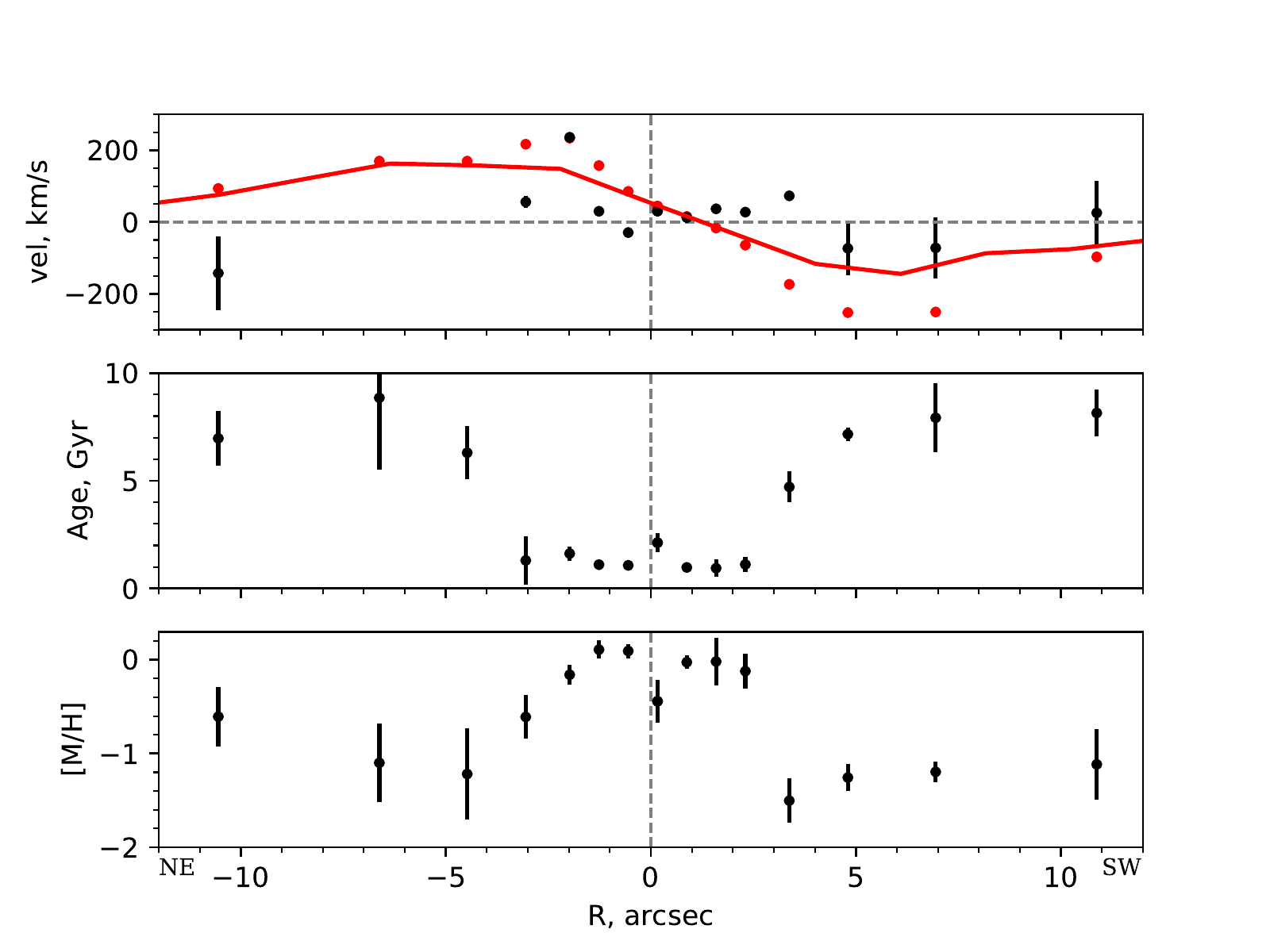}
\caption{Gas/stars kinematics and stellar population properties derived from SCORPIO-2 observations at PA=$60^\circ$. From top to bottom:  line-of-sight velocity distribution for stars (black dots) and ionized gas (red), red line shows the    velocities in this direction according FPI data, the systemic velocity $25482\kms$ was subtracted;   radial distribution of the SSP age; the same for the stellar metallicity.
}
\label{fig:star_properties}
  \end{figure}

Fig.~\ref{fig:star_properties} shows the distribution along the spectrograph's slit of the stellar line-of-sight velocities, age and metallicity derived from the  description of the SCORPIO-2 spectra by the single stellar population (SSP) pPXF model. This plot demonstrates that the inner $r<3-4''$ (5--6.5 kpc) region is decoupled both in kinematic   and   stellar population properties from the outer host galaxy. The  luminosity--averaged age of   stars in  the central region is significantly younger comparing with outer part of the galaxy   ($~\sim1$  vs 7--9 Gyr) and has a near solar metallicity typical for the recent burst of  star formation, whereas [M/H]$\approx-1$ in the outer region that is typical for the ``red-sequence'' early-type galaxy. 

The flux and velocity of the ionized gas emission lines were derived via  Gaussian fitting of galactic  spectra after subtraction of the pPXF model  in each radial bin. During this procedure we accepted the same line-of-sight velocity separately  for the system of the forbidden (\OIII, \NII, \SII) and the Balmer hydrogen lines. The top panel in the  Fig.~\ref{fig:star_properties} demonstrates an  agreement   between velocities of the forbidden lines in the SCORPIO-2 long-slit spectrum (red points) and a pseudo-slit cut through the FPI velocity field (red line) if the differences in the spatial and spectral resolutions will be taken into account.  

\red{According to our Gaussian fitting results the mean flux ratio  of the brightest Balmer lines at $r<12''$ \HaHb=$2.9\pm0.1$ is in a good agreement with the `standard' value of an intrinsic Balmer decrement both for AGN and \HII-regions ($\sim3.1$ and 2.86, see \citet{Groves2012MNRAS.419.1402G} for review).  For this reason, we did not correct the observed spectrum  for interstellar dust extinction. The reddening map presented in \citet{Gagne2014ApJ...792...72G} reveals   that the significant extinction $E(B-V)>0.2$ (possibly related to a circumnuclear dust lane) is detected in the inner $r<2''$. This region is a poorly resolved in our long-slit data with a seeing value $~2.2''$. 
}

The kinematic of stars is significantly differing from  the gaseous component: in the inner $r<5''$ the radial gradient of stellar line-of-sight velocities is near zero with a hint of slow counter-rotation in the SW part of the velocity distribution.  Such kinematic  feature is indicative of a multi-spin galaxy where the gas and stars rotate in  different planes and/or in different directions. Indeed, an absence of a velocity gradient should be observed if we put a slit along the rotating disk minor axis, or if an  unresolved counter-rotating stellar component  presents in the kinematically distinct core  \cite{Young2018MNRAS.477.2741Y,Gasymov2022arXiv220911240G}. Both features are unsurprising after galactic merging or accretion of the  external gas by early type host galaxy \cite{Sil2019ApJS..244....6S}.


\section{The gas ionization}
\label{sec:BPT}

To discriminate the gas excitation mechanism the BPT (after \citep{Baldwin1981PASP...93....5B}) diagnostic diagrams of  the emission line flux  ratios were widely used. In the previous work \cite{Villar2018MNRAS.474.2302V} argued that  even the very external regions of the Teacup giant nebula is illuminated by the AGN ionized radiation. However, the GTC spectra did not cover the green region. It was the reason why  in \cite{Villar2018MNRAS.474.2302V}  a very important   \OIIIHb{} line ratio for the giant EELR was accepted the same with the nuclear value. 

\begin{figure*}
 \centerline{
\includegraphics[height=0.42\textwidth]{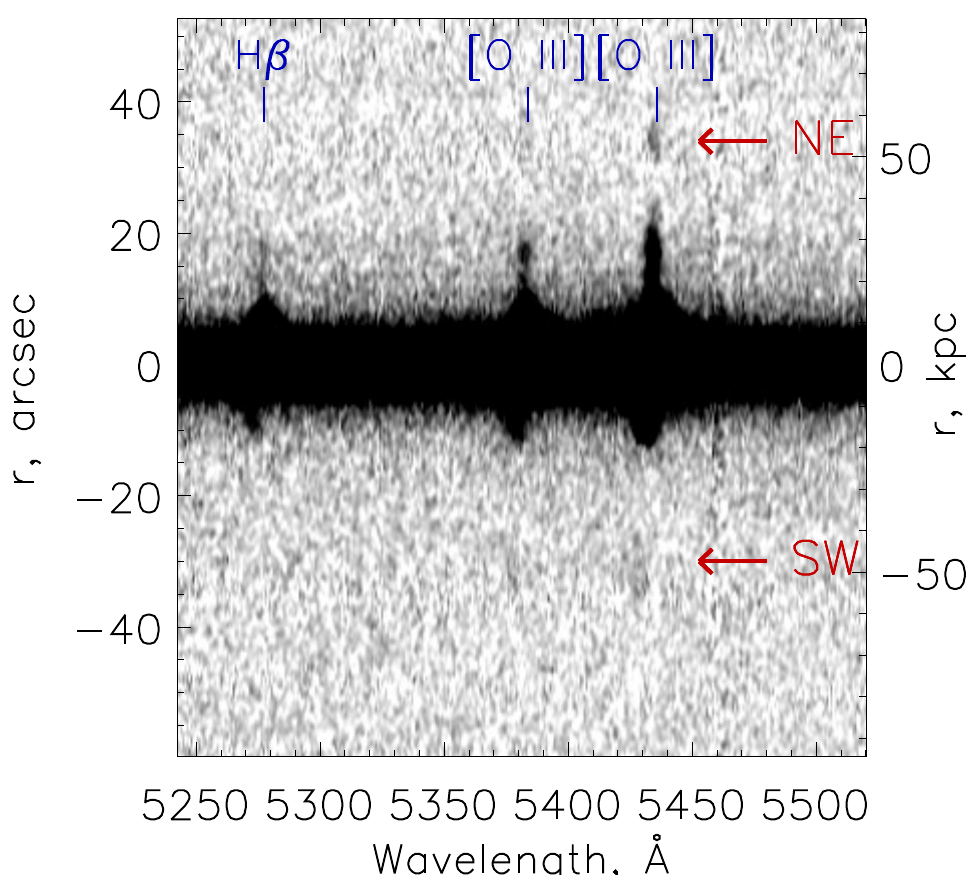}
\includegraphics[height=0.42\textwidth]{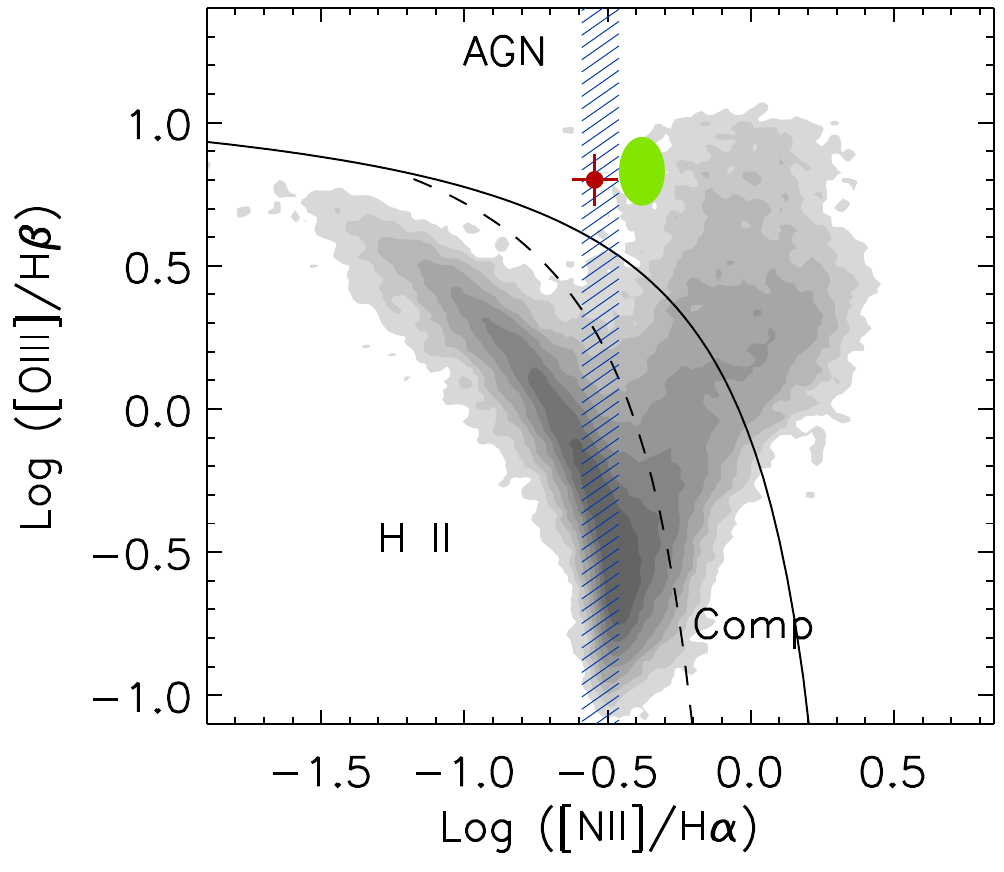}
}
\caption{Ionized gas in the Teacup giant EELR. Left: a fragment of the 2D SCORPIO-2  spectrum along $PA=60^\circ$ around the \Hb+\OIII$\lambda\lambda4959,5007$ region. The red arrows mark the NE and WS Arcs. Right:  BPT diagram similar to one the presented in fig.~3  \citep{Villar2018MNRAS.474.2302V}. \red{The gray contours represent the density distribution of the SDSS galaxies line ratios based on the data from \citet{Kewley2006MNRAS.372..961K}.  The division lines between star-forming galaxies,   AGN and composite nucleus are taken from \citet{Kewley2001ApJ...556..121K,Kauffmann2003MNRAS.346.1055K}.} The vertical blue shadowed area represents the range of values measured for the  giant nebula at GTC \citep{Villar2018MNRAS.474.2302V}.   The green filled ellipse shows the line ratios measured 
by \citet{Gagne2014ApJ...792...72G} for the  nearest to the AGN regions ($r<15$ kpc). The red point  corresponds to  both our SCOPRIO-2 \red{(\OIIIHb)} and GTC  \red{(\NIIHa)}measurements for the distances $r=19-41$ kpc.
}
\label{fig_bpt}
\end{figure*}

In our 6-m telescope spectrum we were able to detect the \OIII{} line in the emission arcs up to $r\approx55$ kpc (fig.~\ref{fig_bpt}, left).  The flux in the \Hb{} was also estimated in the integrated spectrum on  distances $r=12-26''$ (19--41 kpc). This gives the direct estimation of the \OIIIHb=$0.80\pm0.08$ that agrees in the errors with prediction in \cite{Villar2018MNRAS.474.2302V}. The diagram \OIIIHb{} vs \NIIHa{} (fig.~\ref{fig_bpt}, right) clearly demonstrates that the observed line ratios both for the central and external parts of the giant nebula corresponds to AGN-type ionizaion. Here we used \OIIIHb{} and  \NIIHa{} ratios derived from the 6m and GTC observations respectively and accepted that the \NIIHa{} ratio in the NE region $PA=60^\circ$ at $r=12-26''$ is the same as  in the integrated GTC spectral measurements in two apertures at $r=11.8\pm2.4''$ and $16.6\pm1.8''$ according \citet{Villar2018MNRAS.474.2302V}.

The similar situation is in other BPT-diagramms presented in \citet{Villar2018MNRAS.474.2302V}: the line ratios of the emission lines in the  Teacup  EELR  clearly corresponds to the ionization by UV-radiation of  the central QSO.

\section{The extended nebula: morphology and kinematics}
\label{sec:kinematics}

At the first glance, the distribution of the ionized gas derived in SCORPIO-2  observations looks similar with the deep \Ha{} emission line images recently published in \citet{Villar2021}: the whole emission structure up to $r\approx18''$ ($\sim28$~kpc) possesses  the inner bubble ($\sim10$~kpc) and distant emission patches and filaments at $r\approx30-35''$  ($\sim47-55$~kpc). The giant emission nebula is elongated in the same position angle with  the radio jet direction ($PA\approx60^\circ$ \cite{Harrison2015ApJ...800...45H}). However,  the image in the hydrogen recombination line \Ha{} reveals the characteristic Arc+Cavity structure in the NE side of the nebula, and very faint straight filament  in the SW direction. In contrast with this picture, on the image in the high excited forbidden \OIII{}  line (fig.~\ref{fig1}) we see the arc-like structure in the SE side too. It seems to be symmetric with the NE  arc in an agreement with \citet{Villar2021} suggestion that the large scale morphology of the nebula is influenced by AGN.

\begin{figure}
\centerline{
\includegraphics[width=0.48\columnwidth]{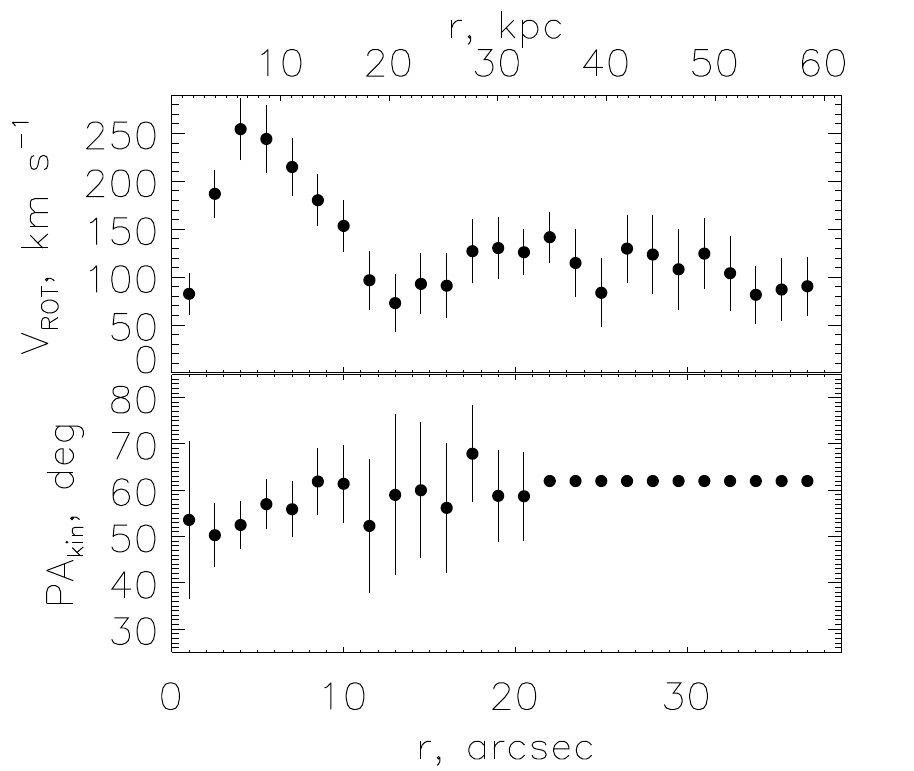}
\includegraphics[width=0.52\columnwidth]{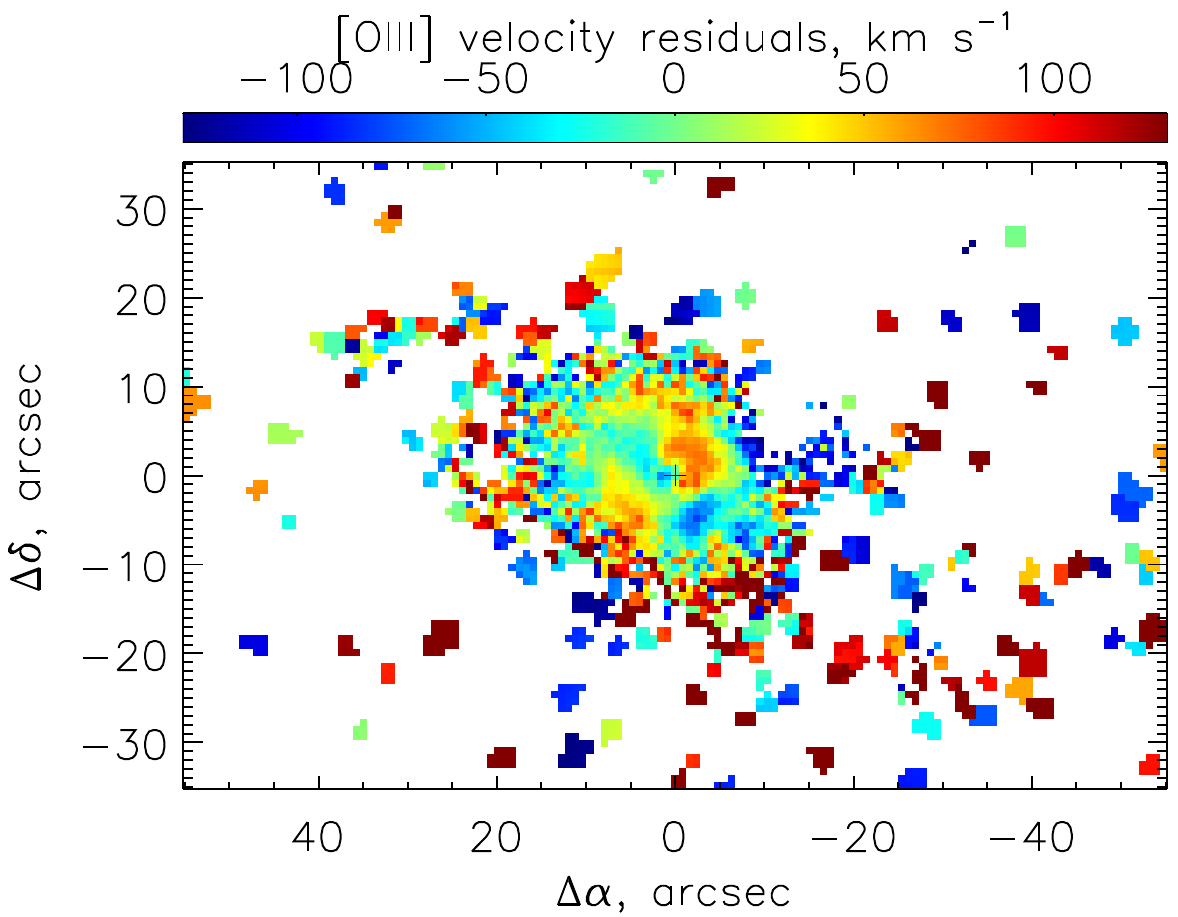}
}
\caption{Radial variations of the tilted-ring model parameters (left panel): the circular
rotation velocity (top) and  the kinematic position angle (bottom). The right panel shows   residual  velocity field after subtraction of the disk circular rotation model.
}
\label{fig:model}
  \end{figure}

We tried to describe the observed line-of-sight ionized gas velocity distribution by the model of a regular circular rotation using our adaptation of a classical `tilted-ring' technique \citep[see][and references therein]{Sil2019ApJS..244....6S,Moiseev2021AstBu..76..316M}. The mean parameters of the gaseous disk orientation were determined from the central  part of the velocity field ($r<25''$): inclination  $i_0=43\pm7^\circ$ and the position angle $PA_0=62\pm4^\circ$.
Fig.~\ref{fig:model} shows the radial variations of the main model parameters: the circular rotation velocity $V_{ROT}$ and kinematic position angle $PA_{kin}$ for the fixed values of the inclination ($i_{kin}=i_0$) and systemic velocity. The $PA_{kin}$ was also  fixed for large radii ($r>21''$) in order to avoid unstable approximation in a sparse velocity field regions.  

It is possible that the rotation curve presented in fig.~\ref{fig:model} is affected by non-circular motions, first of all in the central $r<15$ kpc (location of the giant bubble -- `teacup handle'). This fact is indicated by both the $PA_{kin}$ radial changes and maps of the residual velocities (fig.~\ref{fig:model}, right) with well-ordered patches of negative and positive values. The maximal value of the deviations from the mean rotation ($\pm70\kms$) could be considered as the lower limit on the speed of a large-scale outflow,  the projection effect can increase this value.  Note that the amplitude of these non-circular motions along line-of-sight is   significantly  smaller than $V_{ROT}$ at the corresponded distances.

The  spread of velocity measurements in the faint distant regions prevents us from interpretation of its velocity residuals. Nevertheless, we can conclude that the gas in  both NE and SW arcs lies on the common flat rotation curve  with an amplitude  $100-140\kms$.

\section{Discussion}

There is no doubt that the stellar morphology of the Teacup galaxy reveals the footprint of a merger  event (sec.~\ref{sec:intro}). The observed shell-like narrow tidal  features could be either a result of minor merger (mass ratio $\sim1/10$) of a cold disk system  with a massive elliptical galaxy \cite{Thomson1991MNRAS.253..256T,Shell2008ApJ...684.1062F}, either a moderate merger of two disk galaxies (mass ratio $\sim1/2$, \cite{Shell2022arXiv221016333P}). In the both cases, the lifetime of the observed stellar arcs can be about  1--2 Gyr.

In Sec.~\ref{sec:stars} we find out that the stellar population in the  center of galaxy ($r<5$ kpc)  are significantly younger (SSP age $\sim1$ Gyr) and metal richer ([M/H]$\approx0$) comparing the outer host  (SSP age $\sim7$ Gyr, [M/H]$\approx-1$). These   age of the central starburst put in the above range of the merger age. 
It indicates that the star formation possible triggered by interaction, as it predicted in numerous simulations (\cite{Shell2022arXiv221016333P} and references therein). It is appearing that the same interaction event fed the AGN. In this case, we  do not exclude that the central star formations was partly a result of AGN outflow positive feedback. For instance,   \citep{Keel2015AJ....149..155K} found several candidates for young star clusters in the inner 8 kpc region, outside the ionization cone boundaries. 

The ionized gas distribution agrees with the picture suggested by  \citep{Villar2021} for this and other optically selected Type-2 quasars: in the systems undergoing interaction or merging the ionized gas spread at  over large spatial scales (10--90 kpc) and illuminated by the AGN. Using the long-slit spectroscopy we  directly measured  the indicative line ratio \OIIIHb{} in the external regions  (19-41 kpc) of the Teacup nebula for the first time. Our result fully confirmed the previous suggestion about domination of the AGN ionization  even on these distances \citep{Villar2018MNRAS.474.2302V}. 

The analysis of the \OIII{} velocity field demonstrates that the gas kinematics can be described in the model of a global rotating disk with the line-of-nodes major axis $PA_0=62\pm4^\circ$. This value is significantly differed from the orientation of the stellar continuum isophotes (fig.~\ref{fig1}): $PA_*=157\pm3^\circ$ according to our estimation at  $r=18-20''$. The same orientation of the red continuum isophotes is also appearing in   fig.~A.16 in \cite{Vazdekis2010MNRAS.404.1639V}.  It implies that the gaseous disk is significantly inclined or even orthogonal to the stellar one, that also agrees with  kinematically decoupling stars/gas according to long-slit data (sec.~\ref{sec:stars}):  stellar radial velocity gradient is near zero and significantly smaller than gaseous one, because the slit crosses the stellar host near its minor axis.

The domination of a regular circular rotation in the gaseous kinematics together with the multi-spin gas/stars configuration might constrain  parameters of the interaction produced the Teacup galaxy. It is possible that in the most of the nebula  we are observing not the main galaxy gas, spread by galactic interaction and outflow, but the matter accreted from a companion with a corresponded spin orientation.

The deep \OIII{} emission line image reveals that  the arc-like faint emission structure elongated according to the radio jet direction exists not only in NE direction, as it followed from the previous \Ha{} map by \citet{Villar2021}, but appears also in SW at the same distance from the AGN (50--55 kpc). The shape of the both SW and NE arcs partly repeats the well-known inner emission bubbles, suggesting the possible origin. Could it  be related to a previous AGN outflow?

There are different estimations of current AGN outflow velocities in the region of 10--12 kpc radio bubbles, spanning the range of $V_{out}=50...150\kms$  \citep{Gagne2014ApJ...792...72G,Harrison2015ApJ...800...45H, Ramos2017MNRAS.470..964R}. If we accepted a   conservative  estimation  $V_{out}>70\kms$ according to the \OIII{} residual velocity map (Sec.~\ref{sec:kinematics}), then the dynamical age  of the NE and SW arcs is   $<0.8$~Gyr. This value seems to be in a good agreement with  age  of a central star formation and minor/moderate merging. Therefore, our estimation  does not contradict the fact that the most distant emission arcs  around Teacup  galaxy are related with the first QSO activity episode triggered by galactic interaction that also started a central burst of star formation. \red{Moreover, it is possible that a circumnuclear starburst also contributed to the formation of the external emission arcs via a galactic wind (see \citet{LopezCoba2020AJ....159..167L} for review and observational examples). A more deep study of the external gas ionization properties and its kinematics (including the velocity dispersion distribution)  are needed to separate the possible influence of AGN and starburst-driven wind and radio jet action on a formation of the emission arcs.} 

\section{Conclusion}

3D spectroscopy with scanning FPI is an old  but very powerful technique to study  different astrophysical objects. In this work we present new observational capabilities  of the SAO RAS 6m telescope with the low-resolution FPI that was early used by our team as a tunable filter at 1--2.5m telescopes. The example of the giant nebula related with radio-quiet quasar known as the Teacup galaxy demonstrates that with this device  we are able to map emission lines at the surface brightness level $(few)\times10^{-18}\ergs\,\mbox{cm}^{-2}\,\mbox{arcsec}^{-2}$ during 2 hours of exposures and even study the gas kinematics if the amplitude of velocity changes exceed 50--20$\kms$.  

The Teacup galaxy has been  well studied, including multiwavelength data from X-ray to radio and integral-field spectroscopy in optical and near infrared.  Nevertheless, using  SCORPIO-2 long-slit and 3D spectroscopy we obtained the following new results:

\begin{itemize}
\item The indicative line ratio $log($\OIIIHb{}$)$ was directly estimated  for the  external regions  (19--41 kpc) of the Teacup nebula. \red{The obtained value $0.80\pm0.08$ lies in the range  0.7--0.9 presented  in \citet{Gagne2014ApJ...792...72G}  for the nearest to the AGN region. Together with \NIIHa{} ratio obtained early in GTC observations it 
confirms the domination of AGN radiation in gas ionization in good agreement with conclusion of the paper \cite{Villar2018MNRAS.474.2302V}}.

\item Stars in the inner $r<5$ kpc are significantly younger and metal richer than the outer host galaxy. The starburst age ($\sim1$ Gyr) agrees with  timescale of a merger event \red{proposed in \citet{Keel2015AJ....149..155K}}.

\item The ionized gas velocity field can be described in the term of a circular rotating disk with a flat rotation curve up to distances 50--60 kpc. This disc appears to be significantly inclined or even polar to the stellar host galaxy.

\item The deep map of the \OIII{} emission reveals two symmetric arcs in the external region of the EELR ($r=50-55$ kpc). It might be a remnant of the previous AGN  outflow \red{(may be in a combination with a starburst-driven galactic wind)} with the age $<0.8$ Gyr.
\end{itemize}

An intriguing puzzle is  an alignment of the line-of-nodes of the global rotating gaseous disk and radio jet (and ionization cone) direction.  Whether is this a coincidence or a manifestation of a more powerful AGN influence on the surrounding gas than we expected?
In any case, we hope that the results presented here will be useful for further detailed   simulations of the Teacup system formation, including interaction with a companion and AGN feedback.

\vspace{6pt}

\funding{ This research was funded by  grant No075-15-2022-262 (13.MNPMU.21.0003) of the Ministry of Science and Higher Education of the Russian Federation. }

\acknowledgments{We obtained  the observed data on the unique scientific facility ``Big Telescope Alt-azimuthal''  of SAO RAS. The long-slit observations were peformed by Dmitry Oparin. We thank the  referees (Bobir Toshmatov and Sebastian S{\'a}nchez) and   Cristina Ramos Almeida for their constructive comments  and Aleksandrina Smirnova for her help in preparing the text.   This work is dedicated to the memory  of  Victor   Afanasiev,  whose  enthusiasm  and  work  helped make these observations. Some of the data presented in this paper were obtained from the Mikulski Archive for Space Telescopes (MAST). This research made use of NASA’s
	Astrophysics Data System and the  the NASA/IPAC Extragalactic Database (NED), which is operated by the Jet Propulsion Laboratory, California Institute of Technology, under contract with the National Aeronautics and Space Administration. }

\conflictsofinterest{The authors declare no conflict of interest.} 

\abbreviations{Abbreviations}{
The following abbreviations are used in this manuscript:\\
\noindent 
\begin{tabular}{@{}ll}
ASC & Advanced Camera for Surveys \\
EELR & Extended Emission-Line Regions EELR\\
FPI & Fabry-Perot interferometer\\
GTC & Gran Telescopio Canarias  \\
HST & Hubble Space Telescope \\
SAO RAS & Special Astrophysical Observatory of the Russian Academy of Sciences \\
SSP & Single stellar population \\
\end{tabular}
}

\begin{adjustwidth}{-\extralength}{0cm}

\reftitle{References}

\end{adjustwidth}
\end{document}